\documentclass[epj]{svjour}

\usepackage{bm}

\usepackage{amsfonts}

\usepackage{amscd}
\usepackage{color}
\usepackage{graphicx}

\usepackage{multirow}
\usepackage[colorlinks=true ,citecolor=blue]{hyperref}

 \newcommand{\mmax}{M}
 \newcommand{\eqref}[1]{(\ref{#1})}

\begin{document}
\title{Rashba spin-orbit interaction in graphene armchair nanoribbons}
 \author{Lucia Lenz\inst{1} \and Daniel F. Urban\inst{2} \and Dario Bercioux\inst{3}}

 \institute{Freiburg Institute for Advanced Studies, Albert-Ludwigs-Universit\"at, D-79104 Freiburg, Germany, \email{lucia.lenz@frias.uni-freiburg.de} \and Fraunhofer Institute for Mechanics of Materials IWM, W\"ohlerstra\ss e 11, D-79108 Freiburg, Germany \and Dahlem Center for Complex Quantum Systems and Institut f\"ur Theoretische Physik, Arnimalllee 14, D-14195 Berlin}

\date{\today}
\abstract{We study graphene nanoribbons (GNRs) with armchair edges in the presence of Rashba spin-
orbit interactions (RSOI). We impose the boundary conditions on the tight binding Hamiltonians for bulk
graphene with RSOI by means of a sine transform and study the influence of RSOI on the spectra and
the spin polarization in detail. We derive the low energy approximation of the RSOI Hamiltonian for the
zeroth and first order in momentum and test their ranges of validity. The choice of a basis appropriate for
armchair boundaries is important in the case of mode-coupling effects and leads to results that are easy
to work with. 
}

\PACS{{81.05.ue}{Graphene} \and {75.70.Tj}{Spin-orbit effects} \and {85.75.-d}{Magnetoelectronics} \and {73.63.Bd}{Nanocrystalline materials}}

\maketitle

\section{Introduction}

Spintronics is a multidisciplinary field whose central theme is the active
manipulation of the spin degree-of-freedom in solid state
systems~\cite{fabian:2004}. The integration of spintronics concepts into
standard electronic devices is an important technological challenge for the
semiconductor industry. The goal is the realization of devices, which for equal
sizes, have better performance and lower power consumption with respect to
state-of-the-art electronic devices~\cite{Awschalom:2007}. In this respect,
graphene has great potential for spintronics applications~\cite{Pesin:2012}. On
the one hand it is characterized by very weak spin-orbit interactions
(SOIs)~\cite{Huertas-Hernando2006} making pristine graphene the spintronics
material with the longest spin-coherence time~\cite{Tombros:2007}. On the other
hand, there are several promising proposals for manipulating the spin states via
local material engineering of the graphene
membrane~\cite{Marchenko:2012,Weeks:2011,Barriga2010}.

There are two possible types of SOIs in graphene: intrinsic and  Rashba (R).
Both of them can be understood in  terms of the symmetry properties of the honeycomb
lattice~\cite{Kane2005a} and by tight binding arguments~\cite{Huertas-Hernando2006}.
The intrinsic SOI opens a gap in the energy spectrum and transforms graphene into a
two-dimensional topological insulator~\cite{Kane2005a}.
However, intrinsic SOI is extremely weak in pristine graphene.

The RSOI is mainly connected to the overlap of the $\pi$ and the $\sigma$ orbitals of the carbon atoms in the $sp^2$--hybridization.
It can be tuned by an electric field perpendicular to the graphene plane or by the local curvature of the graphene sheet~\cite{Huertas-Hernando2006}.
Alternatively,  it was proposed that RSOI can be enhanced by modifying the bonding within the graphene sheet, \emph{e.g.} by covering it with hydrogen~\cite{CastroNeto2009a}. To the lowest order, also a rotating magnetic field can induce effects that can be 
considered as an effective RSOI~\cite{Klinovaja2013}.
The predicted RSOI strength for the former two methods is of the order $\lambda\sim 1$meV and approximately one order of magnitude larger for the third method.
In an experimental study~\cite{Marchenko:2012} it was shown that opportune substrate engineering could even lead to $\lambda\sim200$~meV.
Thus there is reason to hope that manipulation of RSOI in graphene will soon allow to realize RSOI based spintronics devices.
In this respect several efforts have already been made in order to investigate the spintronics properties of bulk graphene,
\emph{e.g.} in the case of inhomogeneous SOI structures~\cite{bercioux:2010,bai:2010,lenz:2011,bercioux:2012,Liu2012,marek:2011,Rybkina2013}.

\begin{figure}
  \includegraphics[width=0.95\columnwidth]{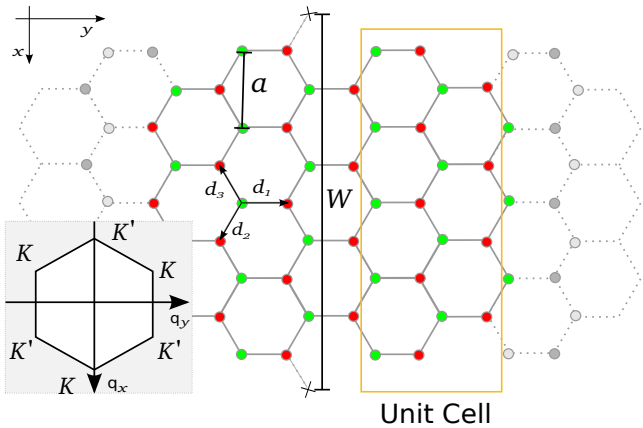}
  \caption{(color online) An armchair GNR of width $\tilde{W}=4 a$, where $a$ is the width of the hexagon.
  The green and red dots mark the A and B atoms of the honeycomb lattice, respectively.
  The distance between the first missing rows of carbon atoms is denoted by $W=\tilde{W}+a$.
  The $\bm{d}_i$ are the displacement vectors connecting nearest neighbor sites.
  The yellow rectangle indicates the unit cell of the ribbon, with $4\times4+2$ atoms for the width shown.
  The inset depicts the first Brillouin zone of bulk graphene with the two inequivalent $K$, $K^\prime$ points.}
  \label{fig:ribbon}
\end{figure}

However, particularly interesting for graphene is the stripe geometry, the so-called graphene nanoribbon (GNR)~\cite{Fuijta1996}.
So far GNRs with RSOI and zig-zag edges have been intensely studied.
For example, Zarea and Sandler~\cite{Zarea2009} analytically derived the energy spectrum of a zig-zag GNR with RSOI produced by an electric
field perpendicular to the GNR surface. Gos\'albez-Mart\'inez \emph{et al.}~\cite{Gosalbez-Martinez2011} studied  RSOI produced by effects of
curvature in zig-zag GNRs. Also GNRs with hardwall boundary conditions and RSOI received some attention~\cite{Stauber2009}. On the contrary,
armchair GNRs with RSOI received little attention so far | despite of their experimental availability.
While large graphene sheets are still polycrystalline and far from perfect, GNRs with an armchair edge (cf. Fig.~\ref{fig:ribbon}) have
already been synthesized with atomic precision and well defined edges~\cite{Cai2010}. In this case, the dangling $\sigma$--bonds of the GNRs
were passivated with hydrogen, so that the $sp^2$--hybridization is preserved also along the edges.

In this article, we fill the gap and study armchair GNRs
with RSOI focusing on the influence of RSOI on the spectrum and the spin polarization. We derive a low energy
theory for the RSOI Hamiltonian starting from the tight
binding approach. Note, that when dealing with effects
that couple different modes, it is important
to start directly from the tight binding
approach and not from an approximative description. \\

The article is organized as follows. Section \ref{sec:method} introduces the system under investigation and reviews some basic theory,
followed by the presentation of our results in Sec.~\ref{sec:AGNRwithRSOI}. The RSOI Hamiltonian for an armchair GNR is derived in
Sec.~\ref{sec:projection} and linearized in order to investigate armchair GNRs in the Dirac approximation in Sec.~\ref{sec:linearization}.
We obtain an analytic equation for the lowest energy bands in Sec.~\ref{sec:anaapprox}, and finally investigate the effect of RSOI on the
spin polarization in Sec.~\ref{sec:spinPolarization}. Specifically, we show that the component of the spin-polarization perpendicular to
the ribbon axis changes sign when the momentum along the axis of the GNR is reversed. This effect arises from the coupling of the GNR bands induced by RSOI.

\section{Basic Theory}
\label{sec:method}
%
%
%
%

\subsection{Armchair graphene nanoribbons}

We study armchair GNRs with the ribbon axis aligned along the $y$-direction as depicted in Fig.~\ref{fig:ribbon}.
The ribbon can be constructed by periodically repeating an unit cell, highlighted by a yellow square, along the $y$-axis.
This unit cell contains four inequivalent atoms for every column of the ribbon and two additional atoms in order to truncate the ribbon in an armchair configuration.
Note that in this article, the distance $W$ between the first \emph{missing} rows of atoms will be used instead of the actual width of the ribbon which is $\tilde{W}=W-a$.

The first Brillouin zone of bulk graphene, which is a hexagon rotated by $\pi/2$ with respect to the graphene lattice, is shown in the inset of Fig.~\ref{fig:ribbon}.
In the absence of RSOI, the conduction and valence bands touch in the two inequivalent Dirac points, $K$ and $K^\prime$, and disperse in a conical shape.


\subsection{Tight binding Hamiltonian}
\label{sec:TBM}
We consider a disorder free system at zero temperature. Our starting point is the Hamiltonian in the tight binding approximation.
The kinetic part restricted to nearest-neighbor hopping is given by~\cite{CastroNeto2009}
%
%
\begin{eqnarray}
  \mathcal{H}_{\mathrm{kin}}&=&-t\sum_{i,j}\sum_{\alpha}[b^\dagger_\alpha(\bm{R}_i+\bm{d}_j)a_\alpha^{\phantom{\dagger}}(\bm{R}_i)+\mathrm{H.c.}]
  \label{eq:tightbindingh0}
\end{eqnarray}
%
%
with the hopping integral $t\approx2.7$~eV.
The operator $a_\alpha(\bm{R}_i)$ annihilates a quasiparticle on the A atom at lattice position $\bm{R}_i$ and with spin $\alpha$
and $b^\dagger_\alpha(\bm{R}_i+\bm{d}_j)$ creates a quasiparticle at the B atom at position $(\bm{R}_i+\bm{d}_j)$ with spin $\alpha$.
The three displacement vectors
%
%
\begin{equation}
\label{eq:displ}
    \bm{d_1}=\frac{a}{\sqrt{3}}(0,1),\quad
    \bm{d_{2/3}}=\frac{a}{\sqrt{3}}\left(\pm\frac{\sqrt{3}}{2},-\frac{1}{2}\right),
\end{equation}
%
%
connect nearest neighbor sites. Here $a=\sqrt{3} a_{\mathrm{cc}}$ is the width of the hexagon and  $a_{\mathrm{cc}}=0.142$~nm the
carbon-carbon distance. In the following, we choose units for which $\hbar\equiv1$. Furthermore, we use $a_{\mathrm{cc}}$ as our unit of length and formally set $a_{\mathrm{cc}}=1$.

The RSOI Hamiltonian induced by, \emph{e.g.} an electric field perpendicular to the graphene sheet, is given by~\cite{Zarea2009,2002,Kuemmeth2009}
%
%
\begin{eqnarray}
  \label{eq:tightbindinghso}
  \mathcal{H}_\mathrm{SO}&=&\mathrm{i} \lambda \sum_{i,j; \alpha,\beta}  b_\alpha^\dag(\bm{R}_i+\bm{d}_j)
  [(\bm{s}\times \bm{\hat{d}}_j)\cdot \bm{\hat{z}}]_{\alpha,\beta} a_\beta(\bm{R}_i)+\mathrm{H.c.} \nonumber \\
\end{eqnarray}
%
%
and has the form of a spin-dependent nearest-neighbor hopping with hopping integral $\lambda$.
Here $\bm{\hat{z}}$ is the unit vector in $z$-direction, $\bm{s}$ is the vector of Pauli matrices associated with
the spin degree of freedom and $\bm{\hat{d}}_j$ are the normalized displacement vectors. The Hamiltonian \eqref{eq:tightbindinghso}
lifts spin degeneracy, since hopping between the A and B atoms is accompanied by a spin-flip.


\subsection{The Dirac approximation}
\label{sec:LWA}

\begin{figure*}
  \centering
  \includegraphics[width=0.8\textwidth]{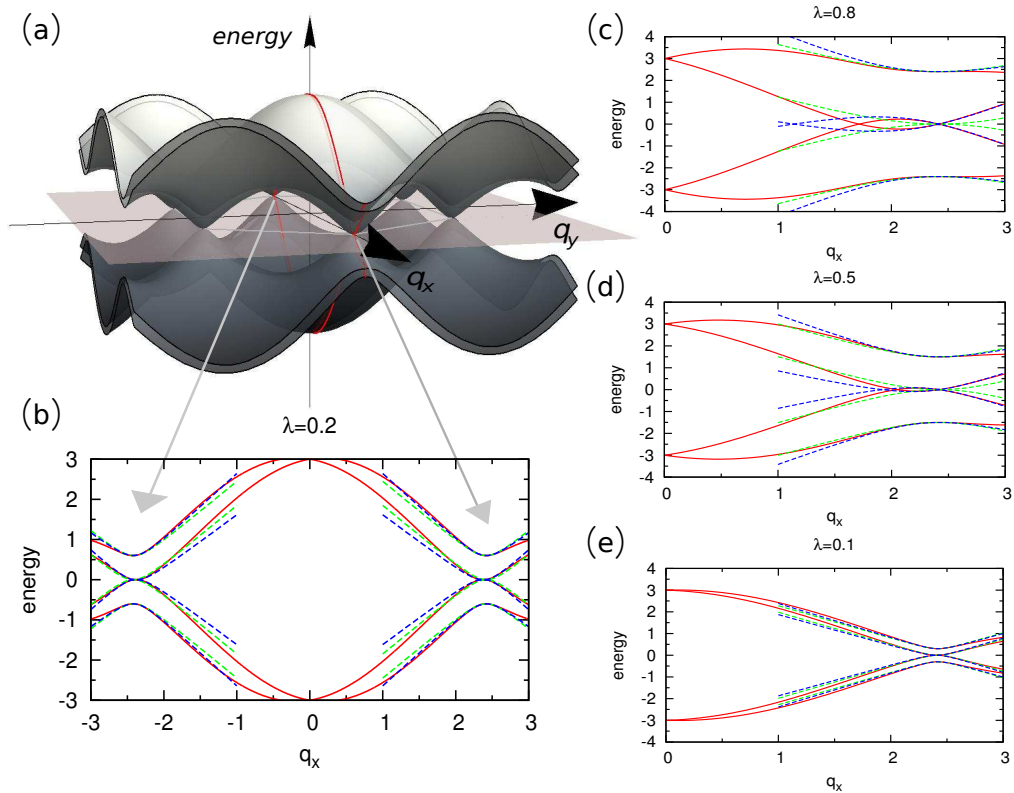}
  \caption{(color online) (a) Energy dispersion of bulk graphene in the presence of RSOI. The cut at constant $q_y=0$, marked in red, is shown in (b).
  The blue dashed line is calculated taking into account $\mathcal{H}_\mathrm{SO}$ up to first order in $k$, for the green dashed line only
  the zeroth order in $k$ is used for $\mathcal{H}_\mathrm{SO}$. (c)-(e) The same cut at $q_y=0$ evaluated for different RSOI strengths $\lambda$.}
  \label{figure:bulk}
\end{figure*}

In order to obtain the Dirac approximation of the nearest-neighbor hopping Hamiltonian \eqref{eq:tightbindingh0} the Fourier transform is taken and expanded close to the Dirac
points $K=(4\pi/(3a),0)=(K_x,0)$ and $K^\prime=-K$~\cite{CastroNeto2009}. This leads to
%
%
\begin{eqnarray}
    \label{eq:ham0}
    \mathcal{H}_{\mathrm{Dirac}}&=&v_\mathrm{\mathrm{F}}(\tau_z s_0 \sigma_x k_x+\tau_0 s_0 \sigma_y k_y),
\end{eqnarray}
%
%
where $k_{x/y}$ are the wavevectors measured with respect to the $K$-points and $v_\mathrm{\mathrm{F}
}=(\sqrt{3} a t)/2$ is the Fermi velocity.
The $\sigma_i$, $s_i$ and $\tau_i$ ($i=x,y,z$) are Pauli matrices representing the A/B-sublattices, the fermionic spin, and the $K$/$K^\prime$
valley degree of freedom, respectively~\footnote{Here and in the following we have set $\hbar=1$.}. The index 0 is used for the unit matrix.
Note that $\mathcal{H}_\mathrm{Dirac}$ contains two blocks $\mathcal{H}_{\mathrm{Dirac}}^K(k_x)=\mathcal{H}_{\mathrm{Dirac}}^{K^\prime}(-k_x)$
related by time-reversal symmetry (TRS).

The RSOI Hamiltonian for bulk graphene in the Dirac approximation up to zeroth order in the wavevector $\bm{k}$~\cite{Huertas-Hernando2006,Kane2005a,Kuemmeth2009},
and with our choice of coordinates reads
%
%
\begin{equation}
  \label{eq:hamso}
  \mathcal{H}^0_\mathrm{SO}= \frac{3\lambda}{2} (\tau_0\sigma_y s_x-\tau_z\sigma_x s_y).
\end{equation}
%
%
As before, the Hamiltonian obeys TRS but now explicitly depends on the spin.

Before we proceed, let us focus on RSOI in bulk graphene for later comparison.
Figure \ref{figure:bulk} shows the energy spectra of bulk graphene for different $\lambda$ values. The spectra obtained from the tight binding
Hamiltonians \eqref{eq:tightbindingh0} and \eqref{eq:tightbindinghso} (solid red lines) are compared to those of the linearized nearest-neighbor
hopping Hamiltonian \eqref{eq:ham0} and the zeroth order RSOI Hamiltonian \eqref{eq:hamso} (dashed green line).
In general there is a good agreement in vicinity of the Dirac points even for moderate $\lambda$. However, for $\lambda\approx0.5t$ or larger,
the occurrence of additional Dirac points is (naturally) not captured. At least the first order expansion of the RSOI Hamiltonian~\cite{Rakyta2010} (dashed blue lines)
needs to be considered in order to captures this effect (referred to as \emph{triangular warping} in bulk graphene) at least qualitatively.

\subsection{Armchair boundary conditions for GNRs without RSOI}
\label{sec:BC}

We briefly review the boundary conditions for armchair GNRs without RSOI in the Dirac approximation~\cite{Brey2006}.
For energies close to the Dirac points the wave function can be written in terms of slowly varying envelope
functions $\phi^K$ and  $\phi^{K^\prime}$. Without taking into account the spin degree of freedom, these functions are two-component spinors
and eigenstates of one of the two equivalent upper blocks or one of the two lower
blocks of equation~\eqref{eq:ham0} in the $(a_K,b_K)$ or $(a_{K^{\prime}},b_{K^{\prime}})$ representation~\cite{CastroNeto2009}.
The wave function on sublattice $\mu=\mathrm{A,B}$ can be written as
%
%
\begin{eqnarray}
  \label{eq:psi}
  \psi_{\mu}(x,y)=\mathrm{e}^{\mathrm{i} K_x x} \phi^K_{\mu}(x,y) + \mathrm{e}^{-\mathrm{i} K_x x} \phi^{K'}_{\mu}(x,y),
\end{eqnarray}
%
%
where $x$ and $y$ are continuous variables. The two functions $\phi^K_{\mu}$ and $\phi_{\mu}^{K'}$ are expansions of the full wave
function at the two different Dirac points $K$ and $K'$ and the fast varying phases $\mathrm{e}^{\pm \mathrm{i} K_x x}$ are required in
order to bring both partial wave functions back into the same coordinate system ($q_x=\pm K_x+k_x$).

For a ribbon that is cut at $x=0$ and $x=W$, all wave functions
on the missing atoms have to be zero (cf. Fig.~\ref{fig:ribbon}). This leads to the boundary conditions
%
%
\begin{eqnarray}
  \label{eq:boundary0}
  0&\!\!=\!\!&\psi_{\mu}(0,y) =\phi^K_{\mu}(0,y)+\phi^{K^\prime}_{\mu}(0,y),
  \\
  \label{eq:boundaryW}
  0&\!\!=\!\!&\psi_{\mu}(W,y) =\mathrm{e}^{\mathrm{i} K_x W}\phi^K_{\mu}(W,y)+\mathrm{e}^{-\mathrm{i} K_x W}\phi^{K^\prime}_{\mu}(W,y).
  \quad
\end{eqnarray}
%
%
The  boundary condition \eqref{eq:boundary0} requires that spinors propagating along the $x$-axis from the $K$ valley are
superimposed with spinors propagating in the opposite direction from the $K'$ valley and \emph{vise versa}, since the corresponding
eigenvectors are equal due to TRS. The boundary condition \eqref{eq:boundaryW} gives the quantization condition on $k_x$,
%
%
\begin{eqnarray}
  \label{eq:qn}
  k_x^{(n)}\equiv k_n = - K_x + \frac{\pi m}{W}=\frac{n\pi}{W} - \frac{4\gamma \pi}{3W}.
\end{eqnarray}
%
%
The integer $m$ ranging from $m=1$ to $\mmax$ labels different transverse \emph{modes} with transversal wavevector
$q_m=\pi m/W$. The total number of modes
$\mmax$ is equal to half the number of atoms in the unit cell, indicated by the yellow rectangle in Fig.~\ref{fig:ribbon}.
For the second equality sign in (\ref{eq:qn}) we rewrite $W=a(3j+\gamma)$ with a positive integer $j$ and $\gamma=-1,0$ or 1.
The modes $n$ range from $- 4 j+1$ to $n_{max}=\mmax- 4 j=4(2j+\gamma)-2$. The shifted modes $n$ are convenient to work with,
because $n=0$ corresponds to the lowest energy bands. The  energy spectrum of the armchair GNR  now reads
\begin{equation}
    E_{n,\epsilon}=\pm v_\mathrm{F}\sqrt{k_y^2+k_n^2},
\end{equation}
Only for $\gamma=0$ the GNR is metallic,
since a zero-energy solution is possible for $n=0$ and $k_y=0$. For $\gamma=\pm1$, the spectrum shows an energy gap equal to $2|q_1|=2\pi/W$.
In the following we will focus on the effect of RSOI on metallic GNRs, though the method employed is general and also applicable if $\gamma\not=0$.

Finally, using the $q_{m}$ as wavevector, the eigenstates of the kinetic Hamiltonian that obey the boundary conditions
\eqref{eq:boundary0} and \eqref{eq:boundaryW} are proportional to sine functions,
%
%
\begin{eqnarray}
  \label{eq:varphi1}
  \Psi_{\epsilon,n}(x,y)\propto \mathrm{e}^{\mathrm{i} k_y y}
  \Phi^K_{k_x,k_y,\epsilon}\sin(q_{m} x)\,.
\end{eqnarray}
%
%
Here, $\Phi^K_{k_x,k_y,\epsilon}$ are the two-component eigenvectors of the Dirac Hamiltonian of the $K$ valley, depending
on the wavevectors of the exponential ansatz and on the sign of the eigenenergy $\epsilon$.

It is important to recall here, that, in order to obtain the low energy approximations~\eqref{eq:ham0} and~\eqref{eq:hamso},
the tight binding Hamiltonians are expanded over a basis set of exponential wave functions. By definition these states do not
fulfill the armchair boundary conditions. In the case of a graphene armchair nanoribbon without RSOI~\cite{Brey2006},
cf. Sec.~\ref{sec:BC}, the standard approach outlined above is to take the plane-wave basis set and then try to satisfy the boundary conditions \emph{a posteriori} by taking appropriate linear combinations.

The more natural choice of basis is to consider sine functions instead, which form a complete basis for a system with armchair boundary conditions.
Both approaches yield the same results as long as terms coupling different modes in the sine transform vanish for the Hamiltonian under consideration -- which indeed is the case for the low energy kinetic Hamiltonian studied in Ref.~\cite{Brey2006}.
On the other hand, the tight binding RSOI Hamiltonian for a GNR couples different modes, and therefore the low energy RSOI Hamiltonian of graphene \eqref{eq:hamso} should not be used, when armchair boundary conditions are studied. In fact, trying to do so leads to non-Hermitian matrices, due to the unbounded derivative operators which appear in the expansion up to 1st order (and higher)~\cite{Rakyta2010} in the momentum of the RSOI Hamiltonian.



\section{Armchair GNRs with RSOI}
\label{sec:AGNRwithRSOI}

\subsection{Sine transform}
\label{sec:projection}

\begin{figure}
  \centering
  \includegraphics[width=1\columnwidth]{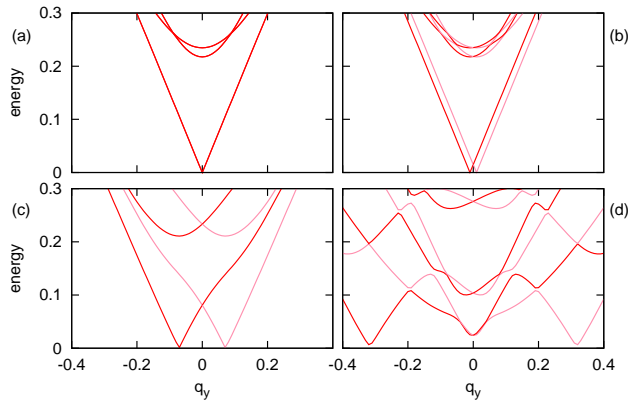}
  \caption{(color online) Energy spectra for a GNR ribbon of width $W = 12a$ and  different strength of the RSOI, (a)
$\lambda=0$, (b) $\lambda= 0.01t$, (c) $\lambda=0.1t$ and (d) $\lambda=0.5t$. Note that an energy gap opens at the Dirac points with increasing $\lambda$. This is analyzed in detail in Fig. 4}
  \label{fig:totspectradifferentW05}
\end{figure}
%

In order to study the effect of RSOI for the case of an armchair GNR we need to expand the Hamiltonians on a
basis set that is naturally fulfilling the boundary condition.
Therefore we take the sine transform of the $a_\alpha(\bm{R}_i)$ and $b_\alpha(\bm{R}_i+\bm{d}_j)$ operators in the tight binding
Hamiltonians \eqref{eq:tightbindingh0} and \eqref{eq:tightbindinghso}.
Taking the sine transform corresponds to choosing a complete set of basis functions $\propto\sin(q_m x)$ with wave vectors $q_m=\pi m/W$.
We use
%

\begin{eqnarray}
 a_\alpha(\bm{R}_i)=\sqrt{\frac{2}{\mmax+1}}\int^\infty_{-\infty}dq_y\mathrm{e}^{\mathrm{i} q_y y_i}\nonumber \\ \sum_{m=1}^{\mmax} \sin(q_m x_i)  a_\alpha(q_m,q_y),\label{eq:a}
\end{eqnarray}
and
\begin{eqnarray}
 b_\alpha(\bm{R}_i+\bm{d}_j)=\sqrt{\frac{2}{\mmax+1}}\int^\infty_{-\infty}dq_y\mathrm{e}^{\mathrm{i} q_y (y_i+ d_{j,y})}\nonumber \\ \sum_{m=1}^{\mmax} \sin\left[q_{m} (x_i+d_{j,x})\right]
 b_\alpha(q_{m},q_y),\label{eq:b}
\end{eqnarray}

%
where $\bm{R}_i=(x_i,y_i)$ and $\bm{d}_{j}=(d_{j,x},d_{j,y})$ .
The ribbon is assumed to be infinite in $y$-direction, therefore a continuous wavevector $q_y$ can be used.
Inserting the transformation \eqref{eq:b} into equation~\eqref{eq:tightbindinghso} yields
%

\begin{eqnarray}
\label{eq:Hsosine}
    \mathcal{H}_\mathrm{SO} &=& \lambda\int dq_y\sum_{m,m'; \alpha=\pm}\left\{\right.
\nonumber\\
    &&\left(\mathcal{A}_{m,m'}^\mathrm{intra}+\alpha\mathcal{A}_{m,m'}^\mathrm{inter}\right)
      b_{-\alpha}^\dag(q_{m},q_y)a_\alpha(q_{m'},q_y)
\nonumber \\
    &&+\left.
    \left(\mathcal{A}_{m',m}^\mathrm{intra}+\alpha\mathcal{A}_{m',m}^\mathrm{inter}\right)^*
    a_{\alpha}^\dag(q_{m},q_y)b_{-\alpha}(q_{m'},q_y)
    \right\}
\nonumber \\
\end{eqnarray}
with
%
%

\begin{eqnarray}
\label{eq:Aintra}
    \mathcal{A}_{m,m'}^\mathrm{intra}=&\mathrm{i}\left[\mathrm{e}^{-\mathrm{i} q_y}-\mathrm{e}^{\mathrm{i} q_y/2}\cos(\sqrt{3}q_m/2)\right]\delta_{m,m'},
\\
\label{eq:Ainter}
    \mathcal{A}_{m,m'}^\mathrm{inter}=&\sqrt{3}\mathrm{e}^{\mathrm{i} \frac{q_y}{2}}\sin(\frac{\sqrt{3}q_m}{2}) \mathcal{B}_{m,m'}\left[\frac{1-(-1)^{m+m'}}{2}\right].
\end{eqnarray}

%
Here $\delta_{m,m'}$ is the Kronecker delta and
%
%
\begin{eqnarray}
    {\mathcal B}_{m,m'}=
    \frac{1}{\mmax+1}\left[\cot\left(\frac{\pi(m+m')}{2(\mmax+1)}\right)\right.\nonumber \\
    \hspace{2cm}\left.-\cot\left(\frac{\pi(m-m')}{2(\mmax+1)}\right)\right].
\end{eqnarray}
%
%
We recall that throughout the paper the carbon-carbon-distance $a_{\rm cc}$ is chosen as the unit of length.
There are two terms in the RSOI Hamiltonian: an intra mode
and an inter mode coupling term. Both couple states with different spin, however, the former is coupling modes with the same mode
index $m=m'$, while the latter couples different modes $m\neq m'$ with different parity | i.e. odd modes with even modes.
The intra mode coupling is a peculiarity of GNRs associated to the two inequivalent carbon atoms in the unit cell.
No similar effect is present in quantum wires with RSOI~\cite{Governale2002,perroni:2007}.

Finally, inserting Eqs.~\eqref{eq:a} and \eqref{eq:b} in $\mathcal{H}_{\mathrm{kin}}$, leads to
%
%
\begin{eqnarray}
 \mathcal{H}_{\mathrm{kin}}=-t\sum_{\alpha=\pm}&\displaystyle\int^\infty_{-\infty}\!\!\!\!dq_y\!\!\sum_{m=1}^{\mmax}\label{eq:H0sine}
\mathrm{e}^{-\mathrm{i} q_y} \left[1+2\mathrm{e}^{\mathrm{i} \frac{3 q_y}{2}}\cos\left(\frac{\sqrt{3}q_m}{2}\right)\right]\nonumber \\ & b^\dagger_\alpha(q_{m},q_y)a_\alpha(q_{m},q_y)+\mathrm{H.c.},
\end{eqnarray}
%
%
where, of course, there is no coupling between different modes.

Figure~\ref{fig:totspectradifferentW05} shows GNR spectra for different RSOI strengths $\lambda$. With increasing
$\lambda$ we first observe a splitting of the bands, as spin degeneracy is lifted. The splitting is increased for larger
$\lambda$ which leads to a deformation of the bands due to avoided band crossings. Moreover, we observe that RSOI opens a small
gap in the energy spectrum. Finally, for even larger $\lambda=0.5t$, we observe that higher mode bands are pressing down
and create an additional minimum in the two lowest bands.
A related creation of additional Dirac points is known from bulk graphene (triangular warping)
\cite{Rakyta2010} and occurs for similar values of SOI strength, cf. Fig. \ref{figure:bulk} c--e. Note, however, that for the ribbon geometry, we observe a spontaneous creation of additional Dirac points at
finite distance from the original ones and not a continuous splitting of the Dirac points, as in bulk graphene.  Due to the coupling between the bands of the GNR induced by RSOI, its dispersion relation can not be related directly to the one of bulk graphene.

\begin{figure}
  \centering
  \includegraphics[width=0.95 \columnwidth]{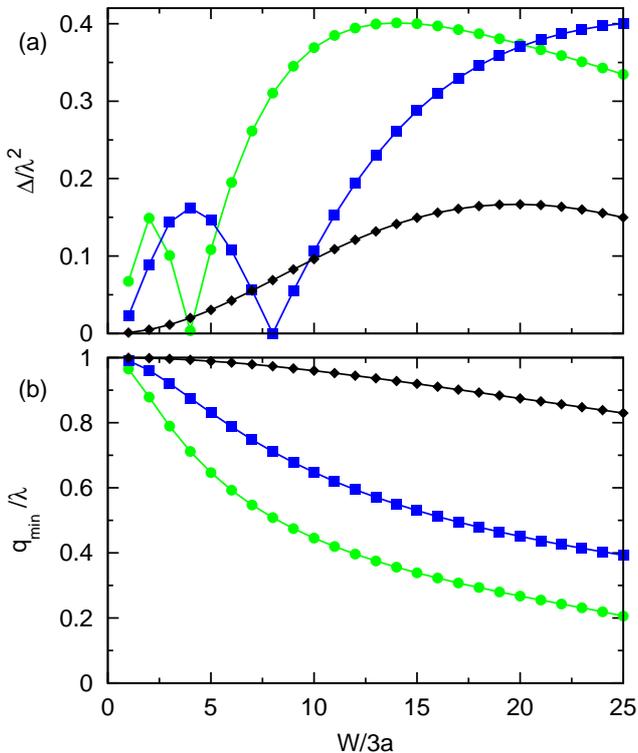}
  \caption{(color online) (a) Energy gap at the Dirac point as a function of the width of the GNR.
  (b) Position of the Dirac point as a function of the width of the GNR.
  For both panels only metallic GNR are considered. Three different strength of the RSOI are shown:
  $\lambda=0.1t$ (green dots), $\lambda=0.05t$ (blue squares) and $\lambda=0.01t$ (black diamonds).
  Symbols are connected as guide to the eye.
  }\label{fig:differentW}
\end{figure}

We note in passing that it is remarkable how strongly the shape of the spectra depends on the widths, especially for strong
RSOI. When the width increases, the bands come closer to each other and
their deformation due to avoided band crossings becomes stronger. In the continuum limit they are merged onto the two spin split bands of bulk graphene.
Fig.~\ref{fig:differentW}  shows how the size of the band gap $\Delta$ and the position of the band minimum change as a function
of the GNR width.  For narrow ribbons, the band minimum is close to $\lambda$ and
moves to smaller values of $q_y$ with increasing width. In the continuum limit we recover
the band minimum at $q_y=0$. The dependency of the band gap on the width has a particularly interesting shape for narrow ribbons. For
larger ribbons the gap goes to zero as expected from bulk graphene.

In the next section we will linearize the Hamiltonians \eqref{eq:Hsosine} and \eqref{eq:H0sine} and investigate for which parameter ranges such a simplified model is valid.

%
%

\subsection{Linearization}
\label{sec:linearization}

\begin{figure}
  \centering
  \includegraphics[width=0.95\columnwidth]{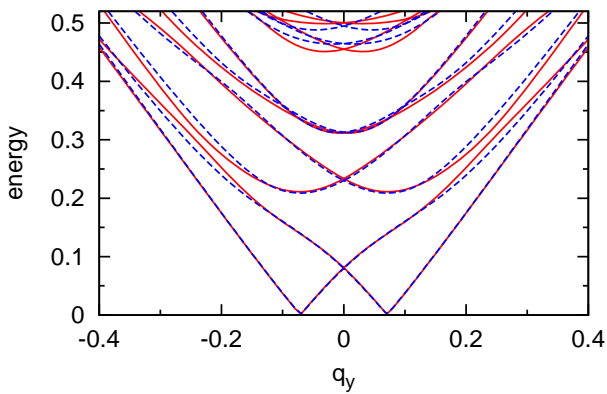}
  \caption{(color online) Energy spectrum for a GNR of width $W=12a$ and RSOI
strength $\lambda=0.1t$. The dashed, blue line corresponds to the linear approximation, the solid red line corresponds to the exact result.}
  \label{fig:linearapprox}
\end{figure}

In this section, we will discuss the linearized versions of the kinetic Hamiltonian (\ref{eq:H0sine}) and the RSOI Hamiltonian (\ref{eq:Hsosine})
with $k_n$ and $k_y$ being the small displacements from the $K$ point,
i.e. $q_m=K_x+k_n$ and $q_y=K_y+k_y$. A Taylor series of
\eqref{eq:H0sine} leads to the well-known Dirac Hamiltonian with quantized wavevector,
%
%
\begin{eqnarray}
  \label{eq:ham0n}
  \left[\mathcal{H}_\mathrm{kin}\right]_n= v_\mathrm{F} s_0 (\sigma_x k_n + \sigma_y k_y),
\end{eqnarray}
%
with $\mathcal{H}_\mathrm{kin}=\int dq_y \sum_n\psi_n^\dag\left[\mathcal{H}_\mathrm{kin}\right]_n\psi_n$
and \\
$\psi_n=\{a_\uparrow(k_n,k_y),b_\uparrow(k_n,k_y),a_\downarrow(k_n,k_y),b_\downarrow(k_n,k_y)\}$.
On the other hand, linearizing the coupling matrices \eqref{eq:Aintra} and \eqref{eq:Ainter} around the $K$-point yields
%
%

\begin{eqnarray}
\label{eq:Aintra:expanded}
    \mathcal{A}_{n,n'}^\mathrm{intra}\simeq&\left(\frac{3\mathrm{i}}{2}+\frac{3(k_y+\mathrm{i} k_n)}{4}\right)\delta_{n,n'},
\\
\label{eq:Ainter:expanded}
    \mathcal{A}_{n,n'}^\mathrm{inter}\simeq&\left(\frac{1-(-1)^{n+n'}}{2}\right)
    \frac{3}{\pi(n'-n)}\times\nonumber \\
    &\left(1-\frac{k_n+k_{n'}}{4}+\frac{\mathrm{i} k_y}{2}
    \right).
\end{eqnarray}

%
%
Now, the linearized version of the intra mode RSOI Hamiltonian \eqref{eq:Hsosine} for modes $n=n'$ reads
%
%
\begin{eqnarray}
\label{eq:Hson}
  \left[\mathcal{H}^\mathrm{intra}_\mathrm{SO}\right]_{n}&=&\frac{3\lambda}{2}s_x
  \left(\sigma_y+\frac{\sigma_xk_y+\sigma_yk_n}{2}\right)
\end{eqnarray}
%
%
with $\mathcal{H}^\mathrm{intra}_\mathrm{SO}=\int dq_y \sum_n\psi_n^\dag\left[\mathcal{H}^\mathrm{intra}_\mathrm{SO}\right]_n\psi_n$.
Comparing the constant terms to the linearized RSOI-Hamiltonian of the infinite graphene plane $\mathcal{H}^0_\mathrm{SO}$,
cf. equation~\eqref{eq:hamso}, we observe that this is the same as adding the blocks for the $K$ and the $K^\prime$ valley and
dividing by $1/2$. This illustrates the coupling of the two valleys induced by the armchair boundary conditions.

For small but different $k_n\not=k_{n'}$ and small $k_y$ the elements of the inter mode RSOI Hamiltonian
are given by
%
%
\begin{eqnarray}
\label{eq:Hsonm}
    \left[\mathcal{H}^\mathrm{inter}_\mathrm{SO}\right]_{n,n'}=&\frac{3\lambda(1-(-1)^{n'+n})}{2\pi\mathrm{i}(n'-n)}s_y \times\nonumber \\
    & \left[\sigma_x+\frac{\sigma_yk_y}{2}-\frac{\sigma_x(k_n+k_{n'})}{4}
    \right]
\end{eqnarray}
where
$
\mathcal{H}_\mathrm{SO}^\mathrm{inter}=\int dq_y \sum_{n,n'}\psi_n^\dag\left[\mathcal{H}^\mathrm{inter}_\mathrm{SO}\right]_{n,n'}\psi_{n'}
$.
The main point about equation~\eqref{eq:Hsonm} is that there is a contribution coupling even and odd bands.

A comparison between the full solution and the linearization (Fig.~\ref{fig:linearapprox}) shows, that the latter
approximation captures the spectrum very well for the first six bands.

The linearized Hamiltonian offers the opportunity to discuss analytically the effects of the mode coupling induced by (\ref{eq:Hsonm}).
We start by considering uncoupled modes. Then the states of mode $n$ are obtained from
$H_n=\left[\mathcal{H}_\mathrm{kin}\right]_n+\left[\mathcal{H}^\mathrm{intra}_\mathrm{SO}\right]_{n}$. From (\ref{eq:ham0n}) and (\ref{eq:Hson}) we infer that the
$\psi_n$ are spin polarized in $x$-direction and there is particle-hole symmetry. Therefore we solve $H_n\psi_{n,s,\varepsilon}=\varepsilon E_{n,s}\psi_{n,s,\varepsilon}$ and obtain
%
%
\begin{eqnarray}
\label{eq:lin0:En}
&  E_{n,s}=v_F\sqrt{\left[k_y+s\frac{\lambda}{t}\left(1+\frac{k_n}{2}\right)\right]^2+\left(k_n+s\frac{\lambda}{t}\frac{k_y}{2}\right)^2} \\
&  \psi_{n,s,\varepsilon}=\left\{s\varepsilon \mathrm{e}^{\mathrm{i}\varphi_{n,s}},s,\varepsilon \mathrm{e}^{\mathrm{i}\varphi_{n,s}},1\right\} \\
&  \varphi_{n,s}=\mathrm{arg}\!\!\left[\left(k_n+s\frac{\lambda}{t}\frac{k_y}{2}\right)-\mathrm{i}\left(k_y+s\frac{\lambda}{t}\left(1+\frac{k_n}{2}\right)\right)\right]
\end{eqnarray}
%
%
where $s=\pm$ is the spin quantum number and $\varepsilon=\pm$ labels the positive/negative energy states. For small $\lambda$ the energies are
\begin{eqnarray}
  E_{n,s}/v_F&\simeq&
      \sqrt{k_y^2+k_n^2}
      +\frac{s\lambda}{t}\frac{k_y(1+k_n)}{\sqrt{k_y^2+k_n^2}}
      +\mathcal{O}(\lambda^2)\quad
\end{eqnarray}
Note that for small RSOI the two states $\psi_{\pm n,s,\varepsilon}$ are almost degenerate.

\begin{figure}
  \centering
  \includegraphics[width=0.95\columnwidth]{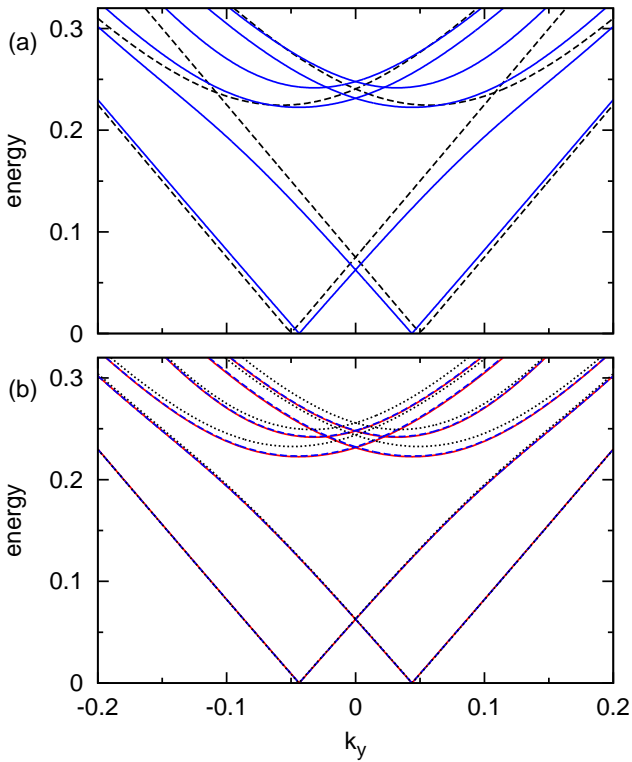}
  \caption{(color online) (a) Energy spectrum as a function of $k_y$ for a GNR of width $W=12a$ and RSOI
  strength $\lambda=0.1t$. The dashed, black lines are obtained with the linear approximation when neglecting the
  inter mode coupling (\ref{eq:Hsonm}).
  The solid blue lines correspond to the linear approximation with the full coupling between the transversal
  modes included.
  (b) Spectrum of the same GNR but obtained from a reduced model including the modes $n=0,\pm1$ (dotted, black lines)
  and $n=0,\pm1,\pm2$ (dashed, blue lines). The latter is in excelent agreement with the exact result (solid red lines) in the energy range shown.
  }
  \label{fig:coupleduncoupled}
\end{figure}

The inter mode coupling (\ref{eq:Hsonm}) has two important effects. First of all, it destroys the $s_x$ spin-polarization since
$\left[\mathcal{H}^\mathrm{inter}_\mathrm{SO}\right]_{n,n'}\propto s_y$. Second, the nearly degeneracy of modes $\psi_{\pm n,s,\varepsilon}$ is
lifted and several avoided band crossings are introduced. This is illustrated in
Fig.~\ref{fig:coupleduncoupled} where the spectrum obtained without considering inter mode coupling is compared to the result from the full linear approximation.
Note that the coupling to the neighboring bands shifts the Dirac Point.

\begin{figure}
  \centering
  \includegraphics[width=1\columnwidth]{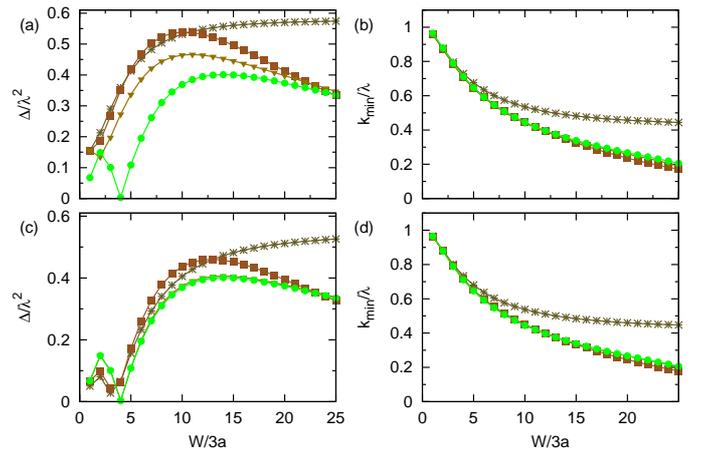}
  \caption{(color online) (a) Bandgap $\Delta$ and (b) position $q_{\rm min}$ of the band minimum as a function of the GNR width obtained from the
  linearized Hamiltonian. The different symbols correspond to considering reduced models containing the modes $n=0,\pm1$ (stars), $n=0,\pm1,\pm2$ (squares),
  and $n=0,\pm1,..\pm10$ (down triangles). The exact width dependence (green dots, c.f. Fig. \ref{fig:differentW}) is shown for comparison.
  (c) and (d) show the bandgap and position of the band minimum obtained from truncating the exact TB Hamiltonian corresponding to the selection of modes given above.}
  \label{fig:figB}
\end{figure}


\subsection{Analytic approximation for the lowest bands}\label{sec:anaapprox}

The currently achievable spin-orbit interactions in physical systems are approximately two orders of magnitude smaller than the
hopping integral $t$~\cite{Marchenko:2012}. Therefore, we study a reduced model for the lowest mode bands, namely $n=0$ ($k_{n=0}=0$) and $n=\pm 1$ ($k_{n=\pm1}=\pm\pi/W$).
We will show, that this is enough to give a good description of the $n=0$ mode, when the inter mode coupling is included. When we are interested in a good approximation for the $n=\pm1$
modes, then we need to include the $n=\pm2$ modes as well.
The most simple approximation for the $n=0$ mode is to consider only $H_0$, and obtain from (\ref{eq:lin0:En}) the close-to-linear energy dispersion
\begin{eqnarray}
  \label{eq:anaapprox}
  E_{0,s} =  v_F\sqrt{(k_y+s\lambda)^2+\frac{k_y^2\lambda^2}{4}}.
\end{eqnarray}
%
%
Here and in the following we set $t\equiv1$.
Equation \eqref{eq:anaapprox} demonstrates that the main correction due to RSOI is the lifting of the spin degeneracy
for the $n=0$ mode bands and a splitting of the Dirac point into two.
Furthermore, there is a term proportional to $k_y^2 \lambda^2$, which opens a gap $\Delta=v_F\lambda^2/\sqrt{4+\lambda^2}$
at the band minimum $k_{\rm min}=\lambda/(1+\lambda^2/4)$ in the spectrum.
Here, both $\Delta$ and $k_{\rm min}$ are independent of the width $W$ which is a crude approximation when comparing to Fig.~\ref{fig:differentW}.
In fact, both quantities are strongly influenced by the coupling with the higher modes which increases with increasing $W$ and $\lambda$.
In the limit of a very wide ribbon $\Delta$ and $k_{\rm min}$ both tend to zero, as expected from bulk graphene.
While including the next higher modes $n=\pm1$ gives a fairly reasonable agreement for the exact width dependence of $k_{\rm min}$,
a large number of modes need to be considered in order to capture the exact width dependence of $\Delta$.
This is illustrated in Fig.~\ref{fig:figB} which compares $\Delta$ and $k_{\rm min}$ obtained from both, the linearized (top panels) and the exact (lower panels) Hamiltonian
when considering a limited number of modes. This comparison also shows that the width dependence is only reproduced qualitatively in the linear approximation for
small and intermediate $W$ and that the oscillating behavior at small $W$ is due to non-linear terms.


\subsection{Spin polarization}
\label{sec:spinPolarization}

Here, we study the spin polarization, that is, the expectation value of the $x$-component of the spin operator for
different subbands
%
%
\begin{equation}
\mathcal{P}_n(q_y)=[\langle S_x=\frac{1}{2}s_x\otimes \sigma_0\rangle]_n\,,
\end{equation}
%
%
as a function of the momentum along the
GNR axis $q_y$ with the tight binding method. In absence of RSOI and external magnetic fields, the spectrum of the GNR
is composed of several sub bands that are two-fold degenerate with opposite spin polarisation. The latter quantity does not depend on the subband index and is constant for all the values of the momentum $q_y$. However, the mixing of the subbands induced by the RSOI modifies the physical picture completely.

\begin{figure}
  \centering
  \includegraphics[width=0.95\columnwidth]{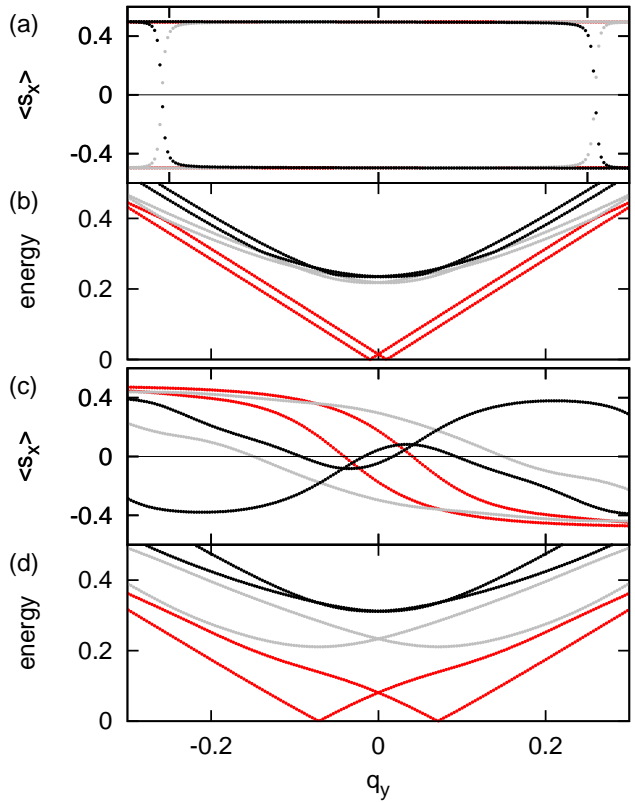}
  \caption{(color online) Spin polarization derived from the tight binding approach for (a) $\lambda=0.01 t$ and (c) $\lambda=0.1
t$
  and a width of $W=12a$. In order to compare the energy bands with the spin polarization, panels (b) and (d) show the
  spectra for the two cases in the same color code: The bands and their polarization are shown in red for $n=0$  and in black and gray for $n=\pm1$.}
  \label{fig:poll001}
\end{figure}

In Figs.~\ref{fig:poll001} (a) and (c) we show the results for two different values of RSOI, $\lambda\sim0.01 t$ and
$\lambda\sim0.1 t$, respectively. Similar to the observations for quantum wires in presence of RSOI,~\cite{Governale2002,perroni:2007}
the spin polarization shows a smooth change to the opposite sign when $q_y$  goes from positive to negative values.
For the lowest energy band ($n=0$), shown in red, the crossover is due to a crossing with a band of the next higher mode ($n=1$),
which is lifted by the terms coupling different modes in $\mathcal{H}_\mathrm{SO}$. The black and gray bands have
the same mode index $n=\pm1$, but still they show a smooth crossover from positive to negative spin polarization, which cannot
be directly attributed to a certain avoided band crossing.
Indeed, in the case of quantum wires with RSOI~\cite{Governale2002,perroni:2007} it is possible to split the system
Hamiltonian in a part with RSOI along the wire direction and another one perpendicular to the first. In this respect,
it is possible to interpret the effects of spin polarization as a function of the mode coupling due to the second term.
This separation of the Hamiltonian in different subparts is not possible in the case of a Dirac-like Hamiltonian.
However, in analogy to the case of quantum wires, we observe that the spin polarization curves do not have a well
defined spin quantization axis. For smaller $\lambda$ the situation is different as then the influence of the higher
mode bands is negligible, since the modes are energetically far apart for ribbons that are narrow enough (c.f. Fig.~\ref{fig:poll001}(a)).
The crossing point of the conduction and the valence band does not influence the spin polarization.
The plots show a clear polarization around $q_y=0$ of either $+1/2$ or $-1/2$ until the first anti-crossing with a higher mode band.

%
%

\section{Summary and discussion}
\label{sec:IV}
We have studied graphene nanoribbons with armchair boundaries and finite Rashba spin-orbit interaction.
The solution of the stationary Schr\"odinger equation requires the choice of a basis set which fulfills the
boundary conditions set by the ribbon geometry. Therefore the momentum representation of the Hamiltonian is
obtained from the real-space tight binding Hamiltonian via a sine transform instead of the (commonly used) Fourier transform.
We emphasize that this procedure is quite different from the case of a graphene armchair nanoribbon without RSOI~\cite{Brey2006}.
There the standard approach is to take a basis set consisting of plane waves, which is obtained via a Fourier transform of the bulk
problem and to apply the boundary conditions \emph{a posteriori} by taking linear combinations of wave function components belonging
to the two different $K$/$K'$ points, cf. equation~\eqref{eq:varphi1}.
But the latter basis cannot be used to diagonalize the RSOI part of the Dirac Hamiltonian \eqref{eq:hamso}. Due to the derivative operators
contained in the expansion up to 1st order in the momentum of the RSOI Hamiltonian~\cite{Rakyta2010}, the use of this standard basis is inconsistent and leads to non-Hermitian matrices.
On the other hand, the boundary conditions are naturally satisfied when using the sine transform.

We have derived the RSOI Hamiltonian in momentum space which allows to study the effects of finite
RSOI on the energy spectrum in a mode dependent manner. Further analytical approximations were presented
for the lowest bands and the case of small RSOI and we have investigated the intra and inter mode coupling induced by RSOI.
We have analyzed the spectra for a variety of ribbon widths and RSOI strengths
For large RSOI the spectrum is strongly influenced
by the width of the ribbon, while for small RSOI this is not the case.
Apart from the overall lifting of spin degeneracy,
the presence of RSOI leads to the
opening of a gap and a shifting of the band minimum. Eventually, sufficiently large RSOI leads to the appearance of an additional Dirac point.
In addition, RSOI also has a significant
effect on the spin polarization perpendicular to the ribbon axis, which depends on the value of the momentum
along the ribbon axis. This effect is due to the mixing of different subbands due to RSOI. The knowledge of
these properties can be useful for  the design of spintronics devices based on graphene nanoribbons.

\begin{acknowledgement} The work of LL and DB is supported by the Excellence Initiative of the German Federal and State Governments and the DFG grant BE 4564/1-1.
We thank Hermann Grabert and Piet Schijven for fruitful discussions.
\end{acknowledgement}


\begin{thebibliography}{99}

\bibitem{fabian:2004} I. Zutic, J. Fabian, and S. das Sarma, Rev. Mod. Phys. \textbf{76}, 323 (2004).

\bibitem{Awschalom:2007} D. D. Awschalom and M. E. Flatte, Nature Physics \textbf{3}, 153 (2007).

\bibitem{Pesin:2012} D.Pesin  and A.H. MacDonald, Nature Mat. \textbf{11}, 409 (2012).

\bibitem{Huertas-Hernando2006} D.~{Huertas-Hernando}, F.~{Guinea}, and A.~{Brataas}.
Phys. Rev. B, \textbf{74}, 155426 (2006).

\bibitem{Tombros:2007} N. Tombros, C. Jozsa, M. Popinciuc, H. T. Jonkman, and B. J. van Wees, Nature \textbf{448}, 571 (2007).


\bibitem{Marchenko:2012} D. Marchenko, A. Varykhalov, M.~R. Scholz, G. Bihlmayer, E.~I.~Rashba, A. Rybkin, A.~M. Shikin, and O. Rader O, Nature Comm. \textbf{3}, 1232 (2012)

\bibitem{Weeks:2011} C. Weeks, J. Hu, J. Alicea, M. Franz, and R. Wu, Phys. Rev. X, \textbf{1}, 021001 (2011).

\bibitem{Barriga2010}
J.~{S{\'a}nchez-Barriga}, A.~{Varykhalov}, M.~R. {Scholz}, O.~{Rader},
  D.~{Marchenko}, A.~{Rybkin}, A.~M. {Shikin}, and E.~{Vescovo},
Diamond and Related Materials, \textbf{19}, 734 (2010).

\bibitem{Kane2005a}
C.~L. {Kane} and E.~J. {Mele}, Phys. Rev. Lett., \textbf{95}, 226801 (2005).




\bibitem{CastroNeto2009a} A.~H. {Castro Neto} and F.~{Guinea},
Phys. Rev. Lett., \textbf{103}, 026804 (2009).
\bibitem{Klinovaja2013} J. Klinovaja and D. Loss, Phys. Rev. X \textbf{3} 011008 (2013)
\bibitem{bercioux:2010}	D. Bercioux and A. de Martino, Phys. Rev. B, \textbf{81}, 165410 (2010).

\bibitem{bai:2010} C. Bai, J. Wang, S. Jia, and Y. Yang, App. Phys. Lett., \textbf{96}, 223102 (2010).

\bibitem{lenz:2011} L. Lenz and D. Bercioux, EPL \textbf{96}, 27006 (2011).

\bibitem{bercioux:2012} D. Bercioux, D.F. Urban, F. Romeo, and R. Citro, Appl. Phys. Lett. \textbf{101}, 122445 (2012).

\bibitem{Liu2012} M.-H. {Liu}, J.~{Bundesmann}, and K.~{Richter}, Phys. Rev. B, \textbf{85}, 085406 (2012).

\bibitem{marek:2011} M. Rataj and J. Barna\'s, Appl. Phys. Lett. \textbf{99}, 162107 (2011).

\bibitem{Rybkina2013} A. A. Rybkina, A. G. Rybkin, V. K. Adamchuk, D. Marchenko, A. Varykhalov, J. S\'anchez-Barriga and A. M. Shikin, ANanotechnology \textbf{24}, 295201 (2013).

\bibitem{Fuijta1996} M. Fujita, K. Wakabayashi K. Nakada and K. Kusakabe,  J. Phys. Soc. Jpn. \textbf{65}, 1902 (1996)


\bibitem{Zarea2009}
M.~{Zarea} and N.~{Sandler}, Phys. Rev. B, \textbf{79}, 165442  (2009).

\bibitem{Gosalbez-Martinez2011}
D.~{Gos{\'a}lbez-Mart{\'{\i}}nez}, J.~J. {Palacios}, and
  J.~{Fern{\'a}ndez-Rossier}, Phys. Rev. B, \textbf{83}, 115436 (2011).

\bibitem{Stauber2009}
T.~{Stauber} and J.~{Schliemann}, New J. Phys., \textbf{11}, 115003 (2009).

\bibitem{Cai2010}
J.~{Cai}, P.~{Ruffieux}, R.~{Jaafar}, M.~{Bieri}, T.~{Braun}, S.~{Blankenburg},
  M.~{Muoth}, A.~P. {Seitsonen}, M.~{Saleh}, X.~{Feng}, K.~{M{\"u}llen}, and
  R.~{Fasel}, Nature, \textbf{466}, 470 (2010).

\bibitem{CastroNeto2009}
A.~H. {Castro Neto}, F.~{Guinea}, N.~M.~R. {Peres}, K.~S. {Novoselov}, and
  A.~K. {Geim}, Rev. Mod. Phys., \textbf{81}, 109 (2009).

\bibitem{Kuemmeth2009} F. Kuemmeth E. I. Rashba, Phys. Rev. B 	\textbf{80} 241409 (2009)
\bibitem{2002}
A.~{de Martino}, R.~{Egger}, K.~{Hallberg}, and C.~A. {Balseiro}, Phys. Rev. Lett., \textbf{88}, 206402 (2002).

\bibitem{Brey2006} L.~{Brey} and H.~A. {Fertig}, Phys. Rev. B, \textbf{73}, 235411 (2006).




\bibitem{Rakyta2010}
P.~{Rakyta}, A.~{Korm{\'a}nyos}, and J.~{Cserti}, Phys. Rev. B, \textbf{82}, 113405 (2010).

\bibitem{Governale2002}
M.~{Governale} and U.~{Z{\"u}licke}, Phys. Rev. B, \textbf{66}, 073311 (2002).

\bibitem{perroni:2007}
C.A. Perroni, D. Bercioux, V. M. Ramaglia, and V. Cataudella,    J. Phys.: Condens. Matter \textbf{19}, 186227 (2007).


\end{thebibliography}
\end{document}